\documentclass[AMA,STIX1COL,doublespace]{WileyNJD-v2}

\usepackage{graphicx}
\usepackage{amsmath}

\usepackage{times}
\usepackage{bm}
\usepackage{enumerate}
\usepackage{xurl}

\usepackage{mathtools}
\usepackage{setspace}
\usepackage{array,booktabs}
\usepackage{tikz}
\usepackage{mathtools}
\usepackage{chngcntr}
\usepackage{algorithm}
\usepackage{algpseudocode}
\usepackage{multirow}
\usepackage{multicol}
\usepackage{subcaption}
\usepackage{amsmath}
\usepackage{footnote}
\usepackage{enumitem}
\usepackage{amsmath}
\usepackage{soul}

\articletype{RESEARCH ARTICLE}%

\received{29 January 2021}
\revised{14 April 2021}
\accepted{27 May 2021}

\begin{document}

\title{An Integrated and Coherent Framework for Point Estimation and Hypothesis Testing with Concurrent Controls in Platform Trials}

\author{Tianyu Zhan, Jane Zhang, Lei Shu and Yihua Gu}

\authormark{Zhan et al.}

\address{Data and Statistical Sciences, AbbVie Inc., North Chicago, Illinois, USA}

\corres{Tianyu Zhan, 1 Waukegan Rd, North Chicago, IL 60064, USA.  \email{tianyu.zhan.stats@gmail.com}}


\abstract[Summary]{A platform trial with a master protocol provides an infrastructure to ethically and efficiently evaluate multiple treatment options in multiple diseases. Given that certain study drugs can enter or exit a platform trial, the randomization ratio is possible to change over time, and this potential modification is not necessarily dependent on accumulating outcomes data. It is recommended that the analysis should account for time periods with different randomization ratios, with possible approaches such as Inverse Probability of Treatment Weighting (IPTW) or a weighted approach by the time period. To guide practical implementation, we specifically investigate the relationship between these two estimators, and further derive an optimal estimator within this class to gain efficacy. Practical guidance is provided on how to construct estimators based on observed data to approximate this unknown weight. {The connection between the proposed method and the weighted least squares is also studied.} We conduct simulation studies to demonstrate that the proposed method can control type I error rate with a reduced estimation bias, and can also achieve satisfactory power and mean squared error (MSE) {with computational efficiency}. Another appealing feature of our framework is the ability to provide consistent conclusions for both point estimation and hypothesis testing. This is critical to the interpretation of clinical trial results. The proposed method is further applied to the Accelerating COVID-19 Therapeutic Interventions and Vaccines (ACTIV) platform trial. }



\maketitle


\section{Introduction}
\label{sec:intro}

A platform trial with a master protocol is able to evaluate multiple treatment options in multiple patient types or diseases within the same overall trial structure.\citep{woodcock2017master} Such an innovative framework can enhance operational efficiency, and make the trial more ethical by utilizing concurrent control.\citep{lu2021practical} Recently, the Food and Drug Administration (FDA) also released a draft guidance to provide recommendations on the design and analysis of trials conducted under a master protocol.\citep{fda2023} This infrastructure is receiving more attention and interest in industry to efficiently meet unmet medical needs.

Despite the aforementioned advantages, a successful platform trial conduct requires much more effort than a traditional clinical trial, with practical considerations discussed in many existing works.\citep{woodcock2017master, lu2021practical, li2022current, li2024unlocking} One unique feature of a platform trial is that the randomization ratio is possible to change over time, when some products enter or exit this trial.\citep{fda2023} Unlike response-adaptive randomization designs or other adaptive clinical trials, this modification is not necessarily dependent on accumulating outcomes data. The FDA guidance recommends that the downstream analysis should account for time periods with different randomization ratios, with possible approaches such as Inverse Probability of Treatment Weighting (IPTW) or a stratified approach by the time period.\citep{fda2023} Section \ref{sec:sim} provides further illustration of the potential negative impact on estimation bias and type I error rate of the traditional method without considering different periods. It then comes with several additional questions for implementation: How to choose a specific analysis strategy? Which method(s) can control the type I error rate and have a reduced estimation bias of the treatment effect? Given a satisfactory performance of type I error rate and bias, which method offers better efficiency? 

To provide some insights into the above questions and to facilitate practical implementation, we investigate two possible approaches discussed in the FDA guidance \cite{fda2023} to accommodate the modification of the randomization ratio in a master protocol, and further propose a coherent strategy for both point estimation and hypothesis testing. The first approach based on Inverse Probability of Treatment Weighting (IPTW) \citep{miguel2020causal} is shown to be a special case of the second option of a weighted method. We also study the choice of the weight parameter to minimize the variance for point estimation, and correspondingly, to increase power for hypothesis testing. Practical guidance on approximating this unknown weight is provided. Another advantage of our framework is its ability to deliver consistent and coherent conclusions for both point estimation and hypothesis testing, as compared with an alternative approach of directly conducting hypothesis testing by the combination testing approach, which is commonly used in adaptive clinical trials to control type I error rate.\citep{bauer1994evaluation, cui1999modification, bretz2009adaptive} This feature is critical to the interpretability and reliability of clinical trial results. {In this work, we specifically focus on the utilization of concurrent controls, which is a more preferred approach to support robust conclusions as compared to non-concurrent controls.} \citep{fda2023}

{There are also some previous works exploring time periods using non-concurrent controls in platform trials.} \citep{lee2020including, bofill2022model, saville2022bayesian, santacatterina2025identification} {In addition to a different scope on concurrent controls, here are some additional contributions of our work. Firstly, our proposed method can accommodate a more general setting with heterogeneous variances across different stages and groups. The typical model-based or regression methods may not have satisfactory performance based on simulation studies in Section} \ref{sec:sim}. {Additionally, we further study the relationship between our proposed method and the weighted regression method to address heterogeneous variance. Our method also demonstrates computational efficiency, as further illustrated in Section} \ref{sec:sim}. 

The remainder of this article is organized as follows. In Section \ref{sec:two}, we discuss these two approaches and our framework in a trial with two stages. Section \ref{sec:multi} generalizes this method to multiple stages. Section \ref{sec:wls} {studies the connection between the proposed method and the weighted least squares.} Simulation studies are conducted in Section \ref{sec:sim} to show that our proposed method has a small bias, accurate type I error rate control with satisfactory power, and consistency in conducting point estimation and hypothesis testing. In Section \ref{sec:real}, we apply this framework to the Accelerating COVID-19 Therapeutic Interventions and Vaccines (ACTIV) platform trial. Discussions are provided in Section \ref{sec:dis}. 

\section{Methods for Two Stages}
\label{sec:two}

In a platform trial, we focus on treatment $\text{T}$ versus placebo $\text{P}$. The randomization ratio between $\text{T}$ and $\text{P}$ is $r_s$, and the number of placebo patients is $n_{s, \text{P}}$, for $s = 1, ..., S$ as the stage indicator. {With a focus on concurrent controls in this work, for the placebo group, we only include subjects who were concurrently randomized to the control and could have been randomized to a treatment drug.} \citep{fda2023} {Taking the Treatment C in Figure} \ref{fig:real} {as an example, placebo data from Stage 2 and Stage 3 are included in the analysis of Treatment C.}

A two-stage trial with $S=2$ is considered for illustration in this section. Section \ref{sec:multi} discusses a generalization to multiple stages. The number of patients randomized to the treatment group $\text{T}$ is therefore $n_{s, \text{T}} = n_{s, \text{P}}\times r_s$ for stage $s$. Suppose that the continuous outcome $y_{s, j}^{(i)}$ of patient $i$ for $i = 1, ..., n_{s, j}$ in stage $s = 1, 2$ and group $j = \text{P}, \text{T}$ follows a Normal distribution,
\begin{equation}
	\label{equ:assump}
	y_{s, j}^{(i)} \sim \mathcal{N}\Big(\mu_{s}+I(j=\text{T})\theta, \sigma^2_{s, j}\Big), 
\end{equation} 
where $\mu_s$ is a common prognostic mean effect for both groups in stage $s$, $\theta$ { is the difference of response means between group T and group P}, and $\sigma_{s, j}$ is the standard deviation for stage $s = 1, 2$ and group $j = \text{P}, \text{T}$. A larger value of $y_{s, j}^{(i)}$ is assumed to be a better outcome. 

Here are some additional discussions of (\ref{equ:assump}):
\begin{enumerate}
\item In the context of this article, $\mu_1$ and $\mu_2$ can be different to reflect time drift in the placebo response, and the ratio $r_1$ and $r_2$ have the flexibility to be distinct. Following the FDA guidance \citep{fda2023}, we focus on a condition that the modification of the $r_2$ from $r_1$ does not depend on accumulating outcomes data. This may occur when certain products enter or exit a platform trial. \citep{fda2023} Section \ref{sec:dis} discusses a generalization with outcome-adaptive features.
\item The placebo response $\mu_s$ in (\ref{equ:assump}) does not change over time within the stage $s$. When determining stages, we can accommodate potential change points of placebo responses in addition to the change of randomization ratios. {To detect such change points or breakpoints, one can implement segmented regression-related approaches, for example, Bayesian change points identification methods.} \citep{carlin1992hierarchical, stephens1994bayesian, lindelov2020mcp} {We conduct an additional analysis to evaluate the robustness of our method when placebo responses are not constant within stage in Section 1 of Supplementary Materials}. Future work is to generalize the current approach to fully address fluctuated or time-varying placebo responses.
\item To facilitate interpretation, we assume that the treatment effect $\theta$ in (\ref{equ:assump}) is invariant across stages.
\end{enumerate}

Our objectives are to provide a point estimate of the treatment effect $\theta$, and to conduct corresponding hypothesis testing of the alternative hypothesis $H_a: \theta>0$ versus the null hypothesis $H_0: \theta \leq 0$ at a significance level of $\alpha$. 

\subsection{A Direct Method}

A direct or traditional method of point estimation is to empirically estimate $\theta$ with pooled data from two groups by $\widehat{\theta}_d$:
\begin{equation}
	\label{equ:direct}
	\widehat{\theta}_d = \frac{\sum_{i = 1}^{n_{1, \text{T}}} y_{1, \text{T}}^{(i)} + \sum_{i = 1}^{n_{2, \text{T}}} y_{2, \text{T}}^{(i)} }{n_{1, \text{T}} + n_{2, \text{T}}} - \frac{\sum_{i = 1}^{n_{1, \text{P}}} y_{1, \text{P}}^{(i)} + \sum_{i = 1}^{n_{2, \text{P}}} y_{2, \text{P}}^{(i)} }{n_{1, \text{P}} + n_{2, \text{P}}}.
\end{equation}
Essentially, the potential impact of different $\mu_s$ in two stages and change of randomization ratios is ignored. 

Consider a statistic $\widehat{z}_d$ defined as,
\begin{equation*}
	\widehat{z}_d = \frac{\widehat{\theta}_d}{\sqrt{\widehat{\sigma}_{\text{T}}^2/(n_{1, \text{T}}+n_{2, \text{T}}) + \widehat{\sigma}_{\text{P}}^2/(n_{1, \text{P}}+n_{2, \text{P}}) }},
\end{equation*}
where $\widehat{\sigma}_{\text{T}}$ and $\widehat{\sigma}_{\text{P}}$ are empirical variance estimators based on $\left[\left\{y_{1, \text{T}}^{(i)} \right\}_{i=1}^{n_{1,\text{T}}}, \left\{y_{2, \text{T}}^{(i)} \right\}_{i=1}^{n_{2,\text{T}}} \right]$ of all treatment data, and $\left[\left\{y_{1, \text{P}}^{(i)} \right\}_{i=1}^{n_{1,\text{P}}}, \left\{y_{2, \text{P}}^{(i)} \right\}_{i=1}^{n_{2,\text{P}}} \right]$ of all placebo data, respectively. The null hypothesis $H_0$ is rejected when $\widehat{z}_d > q_{1-\alpha}$, where $q_{1-\alpha}$ is the upper $\alpha$ quantile of a standard Normal distribution. 

When $\mu_1 = \mu_2$ or $r_1 = r_2$, this direct estimator $\widehat{\theta}_d$ can consistently estimate $\theta$. However, this property is not held under $\mu_1 \ne \mu_2$ and $r_1 \ne r_2$. 

\subsection{Inverse Probability of Treatment Weighting (IPTW)}
\label{sec:IPTW}

To accommodate two stages with different $\mu_s$ and $r_s$, one can use the following estimator $\widehat{\theta}_{IP}$ based on Inverse Probability of Treatment Weighting (IPTW) \citep{miguel2020causal} to estimate $\theta$,
\begin{equation}
	\label{equ:IPTW}
	\widehat{\theta}_{IP} = \frac{\sum_{i = 1}^{n_{1, \text{T}}} y_{1, \text{T}}^{(i)}(r_1+1)/r_1 + \sum_{i = 1}^{n_{2, \text{T}}} y_{2, \text{T}}^{(i)}(r_2+1)/r_2 }{n_{1, \text{T}} + n_{2, \text{T}}+n_{1, \text{P}} + n_{2, \text{P}}} - \frac{\sum_{i = 1}^{n_{1, \text{P}}} y_{1, \text{P}}^{(i)}(r_1+1) + \sum_{i = 1}^{n_{2, \text{P}}} y_{2, \text{P}}^{(i)}(r_2+1) }{n_{1, \text{T}} + n_{2, \text{T}}+n_{1, \text{P}} + n_{2, \text{P}}}.
\end{equation}
The first part estimates the treatment mean if all patients were to receive the treatment with $n_{1, \text{T}} + n_{2, \text{T}}+n_{1, \text{P}} + n_{2, \text{P}}$ in the denominator, and the second part is for the placebo mean. The weight is the inverse probability of receiving treatment, which is $(r_1+1)/r_1$ for $y_{1, \text{T}}^{(i)}$ as an example. 

The hypothesis testing based on $\widehat{\theta}_{IP}$ is discussed in the next section.

\subsection{A Weighted Method}
\label{sec:weight}

Based on the model assumption (\ref{equ:assump}), we can construct the following weighted estimator $\widehat{\theta}_w(w_1)$ based on two stages:
\begin{equation}
	\label{equ:est_w}
	\widehat{\theta}_w(w_1) = w_1\left[\frac{\sum_{i = 1}^{n_{1, \text{T}}} y_{1, \text{T}}^{(i)}  }{n_{1, \text{T}} } - \frac{\sum_{i = 1}^{n_{1, \text{P}}} y_{1, \text{P}}^{(i)} }{n_{1, \text{P}} }\right] + (1-w_1) \left[\frac{\sum_{i = 1}^{n_{2, \text{T}}} y_{2, \text{T}}^{(i)}  }{n_{2, \text{T}} } - \frac{\sum_{i = 1}^{n_{2, \text{P}}} y_{2, \text{P}}^{(i)} }{n_{2, \text{P}} }\right],
\end{equation}
for $w_1 \in [0, 1]$.

It can be shown that $\widehat{\theta}_{IP}$ in (\ref{equ:IPTW}) is equivalent to $\widehat{\theta}_w\left((n_{1, \text{T}} + n_{1, \text{P}})/(n_{1, \text{P}} + n_{1, \text{T}} + n_{2, \text{P}} + n_{2, \text{T}})\right)$, with a weight of $w_1 = (n_{1, \text{T}} + n_{1, \text{P}})/(n_{1, \text{P}} + n_{1, \text{T}} + n_{2, \text{P}} + n_{2, \text{T}})$ proportional to the first stage total sample size:
\begin{align*}
	\widehat{\theta}_{IP} & = \frac{\sum_{i = 1}^{n_{1, \text{T}}} y_{1, \text{T}}^{(i)}(r_1+1)/r_1 + \sum_{i = 1}^{n_{2, \text{T}}} y_{2, \text{T}}^{(i)}(r_2+1)/r_2 }{n_{1, \text{T}} + n_{2, \text{T}}+n_{1, \text{P}} + n_{2, \text{P}}} - \frac{\sum_{i = 1}^{n_{1, \text{P}}} y_{1, \text{P}}^{(i)}(r_1+1) + \sum_{i = 1}^{n_{2, \text{P}}} y_{2, \text{P}}^{(i)}(r_2+1) }{n_{1, \text{T}} + n_{2, \text{T}}+n_{1, \text{P}} + n_{2, \text{P}}} \nonumber \\
	&= \begin{multlined}[t]
		\frac{\sum_{i = 1}^{n_{1, \text{T}}} y_{1, \text{T}}^{(i)}(n_{1, \text{T}} + n_{1, \text{P}})/n_{1,\text{T}} + \sum_{i = 1}^{n_{2, \text{T}}} y_{2, \text{T}}^{(i)}(n_{2, \text{T}} + n_{2, \text{P}})/n_{2,\text{T}} }{n_{1, \text{T}} + n_{2, \text{T}}+n_{1, \text{P}} + n_{2, \text{P}}}  \\[2ex]
		- \frac{\sum_{i = 1}^{n_{1, \text{P}}} y_{1, \text{P}}^{(i)}(n_{1, \text{T}} + n_{1, \text{P}})/n_{1,\text{P}} + \sum_{i = 1}^{n_{2, \text{P}}} y_{2, \text{P}}^{(i)}(n_{2, \text{T}} + n_{2, \text{P}})/n_{2,\text{P}} }{n_{1, \text{T}} + n_{2, \text{T}}+n_{1, \text{P}} + n_{2, \text{P}}}
	\end{multlined}
	\\ 
	&= \begin{multlined}[t]
		\frac{n_{1, \text{T}} + n_{1, \text{P}}}{n_{1, \text{P}} + n_{1, \text{T}} + n_{2, \text{P}} + n_{2, \text{T}}} \left[\frac{\sum_{i = 1}^{n_{1, \text{T}}} y_{1, \text{T}}^{(i)}  }{n_{1, \text{T}} } - \frac{\sum_{i = 1}^{n_{1, \text{P}}} y_{1, \text{P}}^{(i)} }{n_{1, \text{P}} }\right] \\[2ex]
		+ \frac{n_{2, \text{T}} + n_{2, \text{P}}}{n_{1, \text{P}} + n_{1, \text{T}} + n_{2, \text{P}} + n_{2, \text{T}}} \left[\frac{\sum_{i = 1}^{n_{2, \text{T}}} y_{2, \text{T}}^{(i)}  }{n_{2, \text{T}} } - \frac{\sum_{i = 1}^{n_{2, \text{P}}} y_{2, \text{P}}^{(i)} }{n_{2, \text{P}} }\right]
	\end{multlined} \\
	& = \widehat{\theta}_w\left( \frac{n_{1, \text{T}} + n_{1, \text{P}}}{n_{1, \text{P}} + n_{1, \text{T}} + n_{2, \text{P}} + n_{2, \text{T}}}\right). \nonumber
\end{align*}

We further define,
\begin{align*}
	k_1 & = \frac{\sum_{i = 1}^{n_{1, \text{T}}} y_{1, \text{T}}^{(i)}  }{n_{1, \text{T}} } - \frac{\sum_{i = 1}^{n_{1, \text{P}}} y_{1, \text{P}}^{(i)} }{n_{1, \text{P}} },  \\
	k_2 & = \frac{\sum_{i = 1}^{n_{2, \text{T}}} y_{2, \text{T}}^{(i)}  }{n_{2, \text{T}} } - \frac{\sum_{i = 1}^{n_{2, \text{P}}} y_{2, \text{P}}^{(i)} }{n_{2, \text{P}} },  \\
	v_1 & = var\left(k_1\right) = \frac{\sigma^2_{1, \text{T}}}{n_{1, \text{T}}} + \frac{\sigma^2_{1, \text{P}}}{n_{1, \text{P}}}, \\
	v_2 & = var\left(k_2\right) = \frac{\sigma^2_{2, \text{T}}}{n_{2, \text{T}}} + \frac{\sigma^2_{2, \text{P}}}{n_{2, \text{P}}},
\end{align*}
as the two components in (\ref{equ:est_w}), and their corresponding variances, respectively. 

From a point estimation perspective, $\widehat{\theta}_w(w_1)$ in (\ref{equ:est_w}) is a consistent estimator of $\theta$, because it is a weighted average of two stage-wise consistent estimators $k_1$ and $k_2$ with a pre-specified weight $w_1$. \citep{bretz2009adaptive, robertson2023point}

To conduct hypothesis testing based on $\widehat{\theta}_w(w_1)$ including $\widehat{\theta}_{IP}$, we can use the following statistic $\widehat{z}_w(w_1)$,
\begin{equation}
	\label{equ:weight_test}
	\widehat{z}_w(w_1) = \frac{\widehat{\theta}_w(w_1)}{\sqrt{w_1^2 v_1 + (1-w_1)^2 v_2}},
\end{equation}
which has an asymptotic standardized normal distribution, because $k_1$ is independent from $k_2$ given our setup that the modification of randomization ratios does not depend on accumulating outcomes data. The null hypothesis $H_0$ will be rejected if  $\widehat{z}_w(w_1) > q_{1-\alpha}$. Since $\widehat{z}_w(w_1)$ has an asymptotic standardized normal distribution with mean zero under $H_0$, this test has an asymptotic type I error rate control. {Correspondingly, confidence internals of $\widehat{\theta}_w(w_1)$ can be constructed based on this normal approximation.} 

A further question is on how to choose the value of $w_1$ in $\widehat{\theta}_w(w_1)$. The variance of $\widehat{\theta}_w(w_1)$ is
\begin{equation*}
	var\left[\widehat{\theta}_w(w_1) \right] = w_1^2 v_1 + (1-w_1)^2 v_2,
\end{equation*}
By taking a derivative with respect to $w_1$ and setting it as zero, we obtain that 
\begin{equation}
	\label{equ:w_opt_two}
	w_{opt} = \frac{{1}/{v_1}}{{{1}/{v_1} + {1}/{v_2}}}
\end{equation}
is the value minimizing the variance of $\widehat{\theta}_w(w_1)$. This optimal weight $w_{opt}$ for the first stage is interpreted as the proportion of precision (inverse of variance) of data from this stage.   

In practice, $\widehat{\theta}_w(w_{opt})$ cannot be directly used, because $v_1$ and $v_2$ in $w_{opt}$ are unknown quantities. One approach is to use a constant $\widetilde{w}$ based on the best knowledge of variances at the study design stage. Another approach is to use $\widehat{w}$ with $v_1$ and $v_2$ substituted by their empirical estimators, $\widehat{v}_1$ and $\widehat{v}_2$, respectively. Based on the strong consistency of $\widehat{v}_1$ and $\widehat{v}_2$, it can be shown that $\widehat{\theta}_w(\widehat{w})$ is a consistent estimator of $\theta$ and $\widehat{z}_w(\widehat{w}) > q_{1-\alpha}$ is an asymptotic level $\alpha$ test. Simulation studies in Section \ref{sec:sim} evaluate the performance of $\widehat{\theta}_w(\widetilde{w})$ when $\widetilde{w}$ and $w_{opt}$ are different, and the finite sample performance of $\widehat{\theta}_w(\widehat{w})$ with a moderate sample size for late-phase studies. 

\section{Methods for Multiple Stages}
\label{sec:multi}

All results in the previous section for two stages $S=2$ can be directly generalized to a clinical trial with multiple stages ($S>2$), except $w_{opt}$ in $\widehat{\theta}_w(w_{opt})$ with some additional discussion in this section. 

Similar to (\ref{equ:est_w}) in Section \ref{sec:weight}, we denote $\widehat{\theta}_w(\boldsymbol{w})$ as the weighted estimator with $S>2$, 
\begin{equation}
	\label{equ:est_w_multi}
	\widehat{\theta}_w(\boldsymbol{w}) = \sum_{s=1}^S w_s k_s,
\end{equation}
where
\begin{equation*}
	k_s = \frac{\sum_{i = 1}^{n_{s, \text{T}}} y_{s, \text{T}}^{(i)}  }{n_{s, \text{T}} } - \frac{\sum_{i = 1}^{n_{s, \text{P}}} y_{s, \text{P}}^{(i)} }{n_{s, \text{P}} },
\end{equation*}
for $s=1, ..., S$, and $\boldsymbol{w} = (w_1, ..., w_S)$ is a vector of weights with a constraint of $\sum_{s=1}^S w_s = 1$. Denote the variance of $k_s$ as $v_s$. We have the variance of $\widehat{\theta}_w(\boldsymbol{w})$ as
\begin{equation*}
	var\left[\widehat{\theta}_w(\boldsymbol{w})\right] = \sum_{s=1}^S w_s^2 v_s = \sum_{s=1}^{S-1} w_s^2 v_s + \left(1-\sum_{s=1}^{S-1} w_s \right)^2 v_S.
\end{equation*}

To derive a $\boldsymbol{w}$ that minimizes $var\left[\widehat{\theta}_w(\boldsymbol{w})\right]$, we take derivatives of $var\left[\widehat{\theta}_w(\boldsymbol{w})\right]$ with respect to $w_1$, ..., $w_{S-1}$ to obtain the following $S-1$ equations:
\begin{equation*}
	\begin{cases}
		w_1 v_1 - \left(1-\sum_{s=1}^{S-1} w_s \right) v_S = 0, \\
		w_2 v_2 - \left(1-\sum_{s=1}^{S-1} w_s \right) v_S = 0, \\
		... \\
		w_{S-1} v_{S-1} - \left(1-\sum_{s=1}^{S-1} w_s \right) v_S = 0.
	\end{cases}       
\end{equation*}
By solving above equations with the condition of $\sum_{s=1}^S w_s = 1$, we obtain $\boldsymbol{w}_{opt}$ as
\begin{equation}
	\label{equ:w_opt_multi}
	\boldsymbol{w}_{opt} = \left(\frac{1/v_1}{\sum_{s=1}^S 1/v_s}, \frac{1/v_2}{\sum_{s=1}^S 1/v_s},..., \frac{1/v_S}{\sum_{s=1}^S 1/v_s} \right),
\end{equation}
to minimize $var\left[\widehat{\theta}_w(\boldsymbol{w})\right]$. This is consistent with the $w_{opt}$ in (\ref{equ:w_opt_two}) for two stages. 

\section{Weighted Least Squares}
\label{sec:wls}

In this section, we consider regression models to estimate the treatment effect in (\ref{equ:assump}) with two stages $S=2$. To accommodate potentially different $\sigma_{s, j}$ across different stages and different periods, we utilize the weighted least squares (WLS). \citep{montgomery2021introduction}

\subsection{An Oracle Estimator with the Unknown Variance}
\label{sec:wls_oracle}

We first consider an oracle WLS estimator with the unknown variance to compute the weight. Denote the total sample size as $n = n_{1, \text{P}} + n_{1, \text{T}} + n_{2, \text{P}} + n_{2, \text{T}}$. We define the covariate matrix $X$ with a dimension of $n \times 3$, the outcome matrix $\boldsymbol{Y}$ with a dimension $n \times 1$ and the weight matrix $Q$ with a dimension $n \times n$ as following:
\begin{equation}
\label{equ:wls_data}
X = \begin{bmatrix}
\boldsymbol{1}_{1, \text{P}}, & \boldsymbol{0}_{1, \text{P}}, & \boldsymbol{0}_{1, \text{P}}\\
\boldsymbol{1}_{2, \text{P}}, & \boldsymbol{1}_{2, \text{P}}, & \boldsymbol{0}_{2, \text{P}}\\
\boldsymbol{1}_{1, \text{T}}, & \boldsymbol{0}_{1, \text{T}}, & \boldsymbol{1}_{1, \text{T}}\\
\boldsymbol{1}_{2, \text{T}}, & \boldsymbol{1}_{2, \text{T}}, & \boldsymbol{1}_{2, \text{T}}
\end{bmatrix}
\:\:\:\:\:\:\:\:
\boldsymbol{Y} = \begin{bmatrix}
\boldsymbol{y}_{1, \text{P}} \\
\boldsymbol{y}_{2, \text{P}} \\
\boldsymbol{y}_{1, \text{T}} \\
\boldsymbol{y}_{2, \text{T}} 
\end{bmatrix}
\:\:\:\:\:\:\:\:
Q = \begin{bmatrix}
1/{\sigma_{1, \text{P}}^2}, & 0, & \dots & 0 \\
0, & 1/{\sigma_{1, \text{P}}^2}, & \dots & 0 \\
\vdots & \vdots & \ddots & \vdots \\
0, & 0, & \dots & 1/{\sigma_{2, \text{T}}^2}
\end{bmatrix}
\end{equation}
For $s = 1, 2$ and $j = \text{P}, \text{T}$, $\boldsymbol{1}_{s, j}$ is a column vector of $1$, $\boldsymbol{0}_{s, j}$ is a column vector of $0$, and $\boldsymbol{y}_{s, j} = \left[y_{s, j}^{(1)}, \cdots, y_{s, j}^{(n_{s, j})} \right]^T$ is a stack of outcomes. The matrix $X$ has the first column for intercept, the second column as the dummy variable for stage indicator (the second stage $s=2$ is coded as $1$), and the third column as the dummy variable for group indicator (the treatment group $j = \text{T}$ is coded as $1$). The weight matrix $Q$ is a diagonal matrix with $0$ as off-diagonal elements and the corresponding inverse of variance as diagonal elements. Here we treat the variance in $Q$ as known, and consider another method to estimate the variance in the next section. 

Instead of directly conducting weight linear regression on the original data $X$, $\boldsymbol{Y}$ and $Q$, we can implement the simple linear regression on the transformed data $\widetilde{X}$ and $\widetilde{\boldsymbol{Y}}$ to obtain the same estimated coefficients. \citep{montgomery2021introduction} Specifically, $\widetilde{X}$ and $\widetilde{\boldsymbol{Y}}$ are formulated as:
\begin{equation*}
\widetilde{X} = \begin{bmatrix}
1/\sigma_{1, \text{P}}\boldsymbol{1}_{1, \text{P}}, & \boldsymbol{0}_{1, \text{P}}, & \boldsymbol{0}_{1, \text{P}}\\
1/\sigma_{2, \text{P}}\boldsymbol{1}_{2, \text{P}}, & 1/\sigma_{2, \text{P}}\boldsymbol{1}_{2, \text{P}}, & \boldsymbol{0}_{2, \text{P}}\\
1/\sigma_{1, \text{T}}\boldsymbol{1}_{1, \text{T}}, & \boldsymbol{0}_{1, \text{T}}, & 1/\sigma_{1, \text{T}}\boldsymbol{1}_{1, \text{T}}\\
1/\sigma_{2, \text{T}}\boldsymbol{1}_{2, \text{T}}, & 1/\sigma_{2, \text{T}}\boldsymbol{1}_{2, \text{T}}, & 1/\sigma_{2, \text{T}}\boldsymbol{1}_{2, \text{T}}
\end{bmatrix},
\:\:\:\:\:\:\:\:
\widetilde{\boldsymbol{Y}} = \begin{bmatrix}
1/\sigma_{1, \text{P}}\boldsymbol{y}_{1, \text{P}} \\
1/\sigma_{2, \text{P}}\boldsymbol{y}_{2, \text{P}} \\
1/\sigma_{1, \text{T}}\boldsymbol{y}_{1, \text{T}} \\
1/\sigma_{2, \text{T}}\boldsymbol{y}_{2, \text{T}} 
\end{bmatrix}.
\end{equation*}

The least squares estimate $\widehat{\boldsymbol{\beta}}$ based on data $\widetilde{X}$ and $\widetilde{\boldsymbol{Y}}$ is given by:
\begin{equation}
\widehat{\boldsymbol{\beta}} = (\widetilde{X}^T\widetilde{X})^{-1}\widetilde{X}^T \widetilde{\boldsymbol{Y}}, \label{equ:wls}
\end{equation}
which is a $3 \times 1$ matrix. The third element, denoted as $\widehat{\theta}_{\text{WLS}}(\sigma)$, estimates the treatment effect. This is an oracle estimator with true unknown variance $\sigma^2$. 

Next, we show that $\widehat{\theta}_{\text{WLS}}(\sigma)$ is equivalent to the $\widehat{\theta}_w(w_{opt})$ in Section \ref{sec:weight}. Define $B_1 = n_{1, \text{T}} / \sigma_{1, \text{T}}^2$, $B_2 = n_{1, \text{P}} / \sigma_{1, \text{P}}^2$, $B_3 = n_{2, \text{T}} / \sigma_{2, \text{T}}^2$ and $B_4 = n_{2, \text{P}} / \sigma_{2, \text{P}}^2$. The two components $\widetilde{X}^T\widetilde{X}$ and $\widetilde{X}^T \widetilde{\boldsymbol{Y}}$ are expressed as,
\begin{equation}
\label{equ:wls_xy}
\widetilde{X}^T\widetilde{X} = \begin{bmatrix}
B_1 + B_2 + B_3 + B_4, & B_3 + B_4, & B_1 + B_3 \\
B_3 + B_4, & B_3 + B_4, & B_3 \\
B_1 + B_3, & B_3, & B_1 + B_3
\end{bmatrix},
\:\:\:\:\:\:\:\:
\widetilde{X}^T \widetilde{\boldsymbol{Y}} = \begin{bmatrix}
B_1 \frac{\sum_{i = 1}^{n_{1, \text{T}}} y_{1, \text{T}}^{(i)}  }{n_{1, \text{T}}} + B_2 \frac{\sum_{i = 1}^{n_{1, \text{P}}} y_{1, \text{P}}^{(i)}  }{n_{1, \text{P}}} + B_3 \frac{\sum_{i = 1}^{n_{2, \text{T}}} y_{2, \text{T}}^{(i)}  }{n_{2, \text{T}}} + B_4 \frac{\sum_{i = 1}^{n_{2, \text{P}}} y_{2, \text{P}}^{(i)}  }{n_{2, \text{P}}} \\
B_3 \frac{\sum_{i = 1}^{n_{2, \text{T}}} y_{2, \text{T}}^{(i)}  }{n_{2, \text{T}}} + B_4 \frac{\sum_{i = 1}^{n_{2, \text{P}}} y_{2, \text{P}}^{(i)}  }{n_{2, \text{P}}} \\
B_1 \frac{\sum_{i = 1}^{n_{1, \text{T}}} y_{1, \text{T}}^{(i)}  }{n_{1, \text{T}}} + B_3 \frac{\sum_{i = 1}^{n_{2, \text{T}}} y_{2, \text{T}}^{(i)}  }{n_{2, \text{T}}}
\end{bmatrix}.
\end{equation}

The inverse of $\widetilde{X}^T\widetilde{X}$ is given by,
\begin{equation}
\label{equ:wls_xx}
\left(\widetilde{X}^T\widetilde{X}\right)^{-1} = \frac{1}{\text{det}\left(\widetilde{X}^T\widetilde{X}\right)} \begin{bmatrix}
\dots, & \dots, & \dots \\
\dots, & \dots, & \dots \\
-(B_3 + B_4) B_1, & B_1B_4 - B_2 B_3, & (B_1 + B_2)(B_3 + B_4)
\end{bmatrix},
\end{equation}
where $\text{det}\left(\widetilde{X}^T\widetilde{X}\right) = B_1B_2(B_3 + B_4) + B_3 B_4 (B_1 + B_2)$. We only present the third row of $\left(\widetilde{X}^T\widetilde{X}\right)^{-1}$ in (\ref{equ:wls_xx}), because the parameter of interest $\widehat{\theta}_{\text{WLS}}(\sigma)$ is the third element of $\widehat{\boldsymbol{\beta}}$. 

Finally, we show that $\widehat{\theta}_{\text{WLS}}(\sigma)$ as the third element in $(\widetilde{X}^T\widetilde{X})^{-1}\widetilde{X}^T \widetilde{\boldsymbol{Y}}$ with $\left(\widetilde{X}^T\widetilde{X}\right)^{-1}$ in (\ref{equ:wls_xx}) and $\widetilde{X}^T \widetilde{\boldsymbol{Y}}$ in (\ref{equ:wls_xy}) is equivalent to $\widehat{\theta}_w(w_{opt})$. 

\begin{align*}
\widehat{\theta}_{\text{WLS}}(\sigma) = & \frac{1}{\text{det}\left(\widetilde{X}^T\widetilde{X}\right)} \Bigg[ -(B_3 + B_4) B_1 \left(B_1 \frac{\sum_{i = 1}^{n_{1, \text{T}}} y_{1, \text{T}}^{(i)}  }{n_{1, \text{T}}} + B_2 \frac{\sum_{i = 1}^{n_{1, \text{P}}} y_{1, \text{P}}^{(i)}  }{n_{1, \text{P}}} + B_3 \frac{\sum_{i = 1}^{n_{2, \text{T}}} y_{2, \text{T}}^{(i)}  }{n_{2, \text{T}}} + B_4 \frac{\sum_{i = 1}^{n_{2, \text{P}}} y_{2, \text{P}}^{(i)}  }{n_{2, \text{P}}} \right) + \\
& \left(B_1B_4 - B_2 B_3\right) \left(B_3 \frac{\sum_{i = 1}^{n_{2, \text{T}}} y_{2, \text{T}}^{(i)}  }{n_{2, \text{T}}} + B_4 \frac{\sum_{i = 1}^{n_{2, \text{P}}} y_{2, \text{P}}^{(i)}  }{n_{2, \text{P}}} \right) +  (B_1 + B_2)(B_3 + B_4) \left(B_1 \frac{\sum_{i = 1}^{n_{1, \text{T}}} y_{1, \text{T}}^{(i)}  }{n_{1, \text{T}}} + B_3 \frac{\sum_{i = 1}^{n_{2, \text{T}}} y_{2, \text{T}}^{(i)}  }{n_{2, \text{T}}} \right) \Bigg] \\
= & w_{opt}\left[\frac{\sum_{i = 1}^{n_{1, \text{T}}} y_{1, \text{T}}^{(i)}  }{n_{1, \text{T}} } - \frac{\sum_{i = 1}^{n_{1, \text{P}}} y_{1, \text{P}}^{(i)} }{n_{1, \text{P}} }\right] + (1-w_{opt}) \left[\frac{\sum_{i = 1}^{n_{2, \text{T}}} y_{2, \text{T}}^{(i)}  }{n_{2, \text{T}} } - \frac{\sum_{i = 1}^{n_{2, \text{P}}} y_{2, \text{P}}^{(i)} }{n_{2, \text{P}} }\right] \\
\equiv & \widehat{\theta}_w(w_{opt}). 
\end{align*}

For multiple stages with $S > 2$, we need to expand the second column in $\widetilde{X}$ for $S=2$ to $S-1$ columns for dummy variables to represent $S$ stages. Non-trivial additional work is needed to study the relationship between $\widehat{\theta}_{\text{WLS}}(\sigma)$ and $\widehat{\theta}_w(w_{opt})$. This will be reported in a future technical report. Some empirical simulation results show that the equivalence holds for $S=3$ and $S=4$. 

\subsection{A Practical Estimator with Estimated Variance}
\label{sec:wls_est}

In practice, the variance $\sigma^2_{s, j}$ from different stages and different groups are unknown. We adopt the following typical procedure \citep{montgomery2021introduction, shao2008mathematical} to obtain a practical estimator $\widehat{\theta}_{\text{WLS}}(\widehat{\sigma})$ with $\widehat{\sigma}$ estimated by observed data. 
\begin{itemize}
	\item Step 1: Conduct the simple linear regression on data $X$ and $\boldsymbol{Y}$ in (\ref{equ:wls_data}). 
	\item Step 2: For each stage $s$ and group $j$, obtain its estimated variance $\widehat{\sigma}^2_{s, j}$ by the mean of squared residuals from the same stage and the same group. It can be shown that $\widehat{\sigma}^2_{s, j}$ is a consistent estimator of ${\sigma}^2_{s, j}$. \citep{shao2008mathematical} The unknown matrix $Q$ in (\ref{equ:wls_data}) is estimated by $\widehat{Q}$ with $\widehat{\sigma}^2$.
	\item Step 3: Conduct the weighted linear regression in Section \ref{sec:wls_oracle} with the estimated weight matrix $\widehat{Q}$ to obtain $\widehat{\theta}_{\text{WLS}}(\widehat{\sigma})$. 
\end{itemize}

The performance of $\widehat{\theta}_{\text{WLS}}(\widehat{\sigma})$ and $\widehat{\theta}_w(\widehat{w})$ in Section \ref{sec:weight} are evaluated by simulation studies in Section \ref{sec:sim}. 

\section{Simulation Studies}
\label{sec:sim}

In this section, we perform simulation studies with two stages $S=2$ to evaluate the performance of those estimators based on the following setup: $\mu_1 = 0$, $r_1 = 1$, $n_{1, \text{P}} = n_{2, \text{P}} = 120$. The next section considers a multiple-stage setup. There are 7 different scenarios considered:
\begin{itemize}
	\item S1: $\mu_2 = 0$, $r_2 = 1$, $\sigma_{1, \text{P}} = \sigma_{2, \text{P}} = 2$, $\sigma_{1, \text{T}} = \sigma_{2, \text{T}} = 2$
	\item S2: $\mu_2 = 0$, $r_2 = 0.5$, $\sigma_{1, \text{P}} = \sigma_{2, \text{P}} = 2$, $\sigma_{1, \text{T}} = \sigma_{2, \text{T}} = 2$
	\item S3: $\mu_2 = 0.3$, $r_2 = 1$, $\sigma_{1, \text{P}} = \sigma_{2, \text{P}} = 2$, $\sigma_{1, \text{T}} = \sigma_{2, \text{T}} = 2$
	\item S4: $\mu_2 = 0.3$, $r_2 = 0.5$, $\sigma_{1, \text{P}} = \sigma_{2, \text{P}} = 2$, $\sigma_{1, \text{T}} = \sigma_{2, \text{T}} = 2$
	\item S5: $\mu_2 = 0.3$, $r_2 = 0.5$, $\sigma_{1, \text{P}} = \sigma_{2, \text{P}} = 1$, $\sigma_{1, \text{T}} = \sigma_{2, \text{T}} = 4$
	\item S6: $\mu_2 = 0.3$, $r_2 = 0.5$, $\sigma_{1, \text{P}} = \sigma_{2, \text{P}} = 4$, $\sigma_{1, \text{T}} = \sigma_{2, \text{T}} = 1$
	\item S7: $\mu_2 = 0.3$, $r_2 = 0.5$, $\sigma_{1, \text{P}} = 1$, $\sigma_{2, \text{P}} = 2$, $\sigma_{1, \text{T}} = 2$, $\sigma_{2, \text{T}} = 3$
\end{itemize}
S1 corresponds to a basic setting with equal placebo means ($\mu_1 = \mu_2$) and equal randomization ratios ($r_1 = r_2$) of two stages, and homogeneity of variance between two groups. S2 is a variant with $r_2 = 0.5$ and S3 has $\mu_2 = 0.3$. S4 to S6 consider both $r_2 = 0.5$ and $\mu_2 = 0.3$ with different standard deviation assumptions. S7 is more general with different standard deviations across stages. For each scenario, $\theta$ is evaluated at $0$ and a positive value to evaluate type I error rate, and power, respectively. 

We evaluate the following estimators: a direct estimator $\widehat{\theta}_d$ in (\ref{equ:direct}), $\widehat{\theta}_{IP}$ in (\ref{equ:IPTW}) or equivalently $\widehat{\theta}_w\left[(n_{1, \text{T}} + n_{1, \text{P}})/(n_{1, \text{P}} + n_{1, \text{T}} + n_{2, \text{P}} + n_{2, \text{T}})\right]$, $\widehat{\theta}_w(\widetilde{w})$ in (\ref{equ:est_w}) with the optimal weight $w_{opt}$ in (\ref{equ:w_opt_two}) calculated by an assumption of equal variance of two groups, $\widehat{\theta}_w(\widehat{w})$ with $\widehat{w}$ calculated by data to approximate $w_{opt}$, an oracle estimator $\widehat{\theta}_w(w_{opt})$ with the unknown optimal weight $w_{opt}$, {an oracle WLS estimator} $\widehat{\theta}_{\text{WLS}}({\sigma})$ {with unknown variance in Section} \ref{sec:wls_oracle}, {the least squares estimator} $\widehat{\theta}_{\text{LS}}$ {as a direct application of model-based approaches} \citep{lee2020including, bofill2022model, saville2022bayesian, santacatterina2025identification}, {and the WLS estimator} $\widehat{\theta}_{\text{WLS}}(\widehat{\sigma})$ {with estimated variance in Section} \ref{sec:wls_est}. Two oracle estimators $\widehat{\theta}_w(w_{opt})$ and $\widehat{\theta}_{\text{WLS}}({\sigma})$ are included as a reference, but cannot be directly used in practice. All simulation results in this article are based on $10^6$ simulation iterations, unless specified otherwise. 

Table \ref{tab:sim_1_bias} presents the bias and mean squared error (MSE) of those parameters. In S1-S3, all estimators {have numerically small biases that are close to zero} and similar MSEs, except that $\widehat{\theta}_d$ has the smallest MSE under S2. {To characterize simulation errors, we report the mean and standard error of bias of} $\widehat{\theta}_w(\widehat{w})$ {across 10 rounds} ($10^6$ {iterations per round) under S1 as} $2.3\times 10^{-5}$ and $2.0 \times 10^{-4}$, respectively. For S4-S7 with different $\mu_1$ and $\mu_2$, and different $r_1$ and $r_2$, the traditional estimator $\widehat{\theta}_d$ has a non-negligible bias, but the other estimators can accurately estimate the treatment effect $\theta$. In terms of MSE, $\widehat{\theta}_w(\widehat{w})$ has a similar performance with the oracle estimator $\widehat{\theta}_w(w_{opt})$. Under S4 with equal-variance, $\widehat{\theta}_w(\widetilde{w})$ is identical to $\widehat{\theta}_w(w_{opt})$, and therefore has a small MSE. This estimator $\widehat{\theta}_w(\widetilde{w})$ under S5-S7, and the IPTW estimator $\widehat{\theta}_{IP}$ under S4-S7 have slightly increased MSE, but are still comparable with $\widehat{\theta}_w(w_{opt})$. {When it comes to the comparison between the proposed weighted estimators and regression methods, both oracle estimators} $\widehat{\theta}_w(w_{opt})$ and $\widehat{\theta}_{\text{WLS}}({\sigma})$ {have exactly the same numerical results within a scenario. This is consistent with the derivation of their equivalence in Section} \ref{sec:wls_oracle}. {The typical least squares estimator} $\widehat{\theta}_{\text{LS}}$ {can have an increased MSE, e.g., under S7. The practical WLS estimator} $\widehat{\theta}_{\text{WLS}}(\widehat{\sigma})$ {has a very similar performance with the weighted estimator} $\widehat{\theta}_w(\widehat{w})$. {Our proposed estimator} $\widehat{\theta}_w(\widehat{w})$ {is more computationally efficient with approximately} $15$ {minutes to simulate results in all} $7$ {scenarios, as compared to}  $\widehat{\theta}_{\text{WLS}}(\widehat{\sigma})$ with $231$ minutes.

For type I error rate and power in Table \ref{tab:sim_1_power}, observations are similar to the bias and MSE discussed above. The traditional estimator $\widehat{\theta}_d$ has an accurate type I error rate at the nominal level $\alpha=0.05$ under S1-S3, but not S4-S7. The typical least squares estimator $\widehat{\theta}_{\text{LS}}$ has an inaccurate type I error rate under S5-S7. The other estimators can control the type I error rate under all scenarios. Under S2, $\widehat{\theta}_d$ has the largest power, which is consistent with the smallest MSE reported in Table \ref{tab:sim_1_bias}. Under S4-S7, $\widehat{\theta}_w(\widehat{w})$ has a similar power with the oracle estimator $\widehat{\theta}_w(w_{opt})$, while $\widehat{\theta}_{IP}$ and $\widehat{\theta}_w(\widetilde{w})$ have slightly decreased power. The power of $\widehat{\theta}_{\text{LS}}$ under S5 and S7 is not shown due to its type I error rate inflation. Both $\widehat{\theta}_d$ and $\widehat{\theta}_{\text{LS}}$ have power loss under scenarios with deflated type I error rate. {Similar to the previous paragraph on bias and MSE}, $\widehat{\theta}_w(\widehat{w})$ {is more computationally efficient than} $\widehat{\theta}_{\text{WLS}}(\widehat{\sigma})$. {They have numerically similar but not exactly the same performance, because the inference of} $\widehat{\theta}_w(\widehat{w})$ {is based on Normal approximation in Section} \ref{sec:weight} while $\widehat{\theta}_{\text{WLS}}(\widehat{\sigma})$ {is based on the t-distribution. In Section 2 of the Supplementary Materials, we conduct additional analysis with $S=3$ stage with consistent findings. }

From a study design perspective, it is also of great interest to compute planned or target power based on certain design parameters. The asymptotic normality of the weighted estimator in (\ref{equ:est_w}) facilitates such computation. Table \ref{tab:sim_1_pow} shows that the planned power based on $w_{opt}$ and design parameters are close to the empirical power of $\widehat{\theta}_w(w_{opt})$ and $\widehat{\theta}_w(\widetilde{w})$ in each scenario. On the other hand, if the target power is known, we can calculate the sample size based on certain design parameters, including randomization ratios. 

\begin{figure}
	\centering
	\begin{subfigure}[b]{0.55\textwidth}
		\includegraphics[width=1\linewidth]{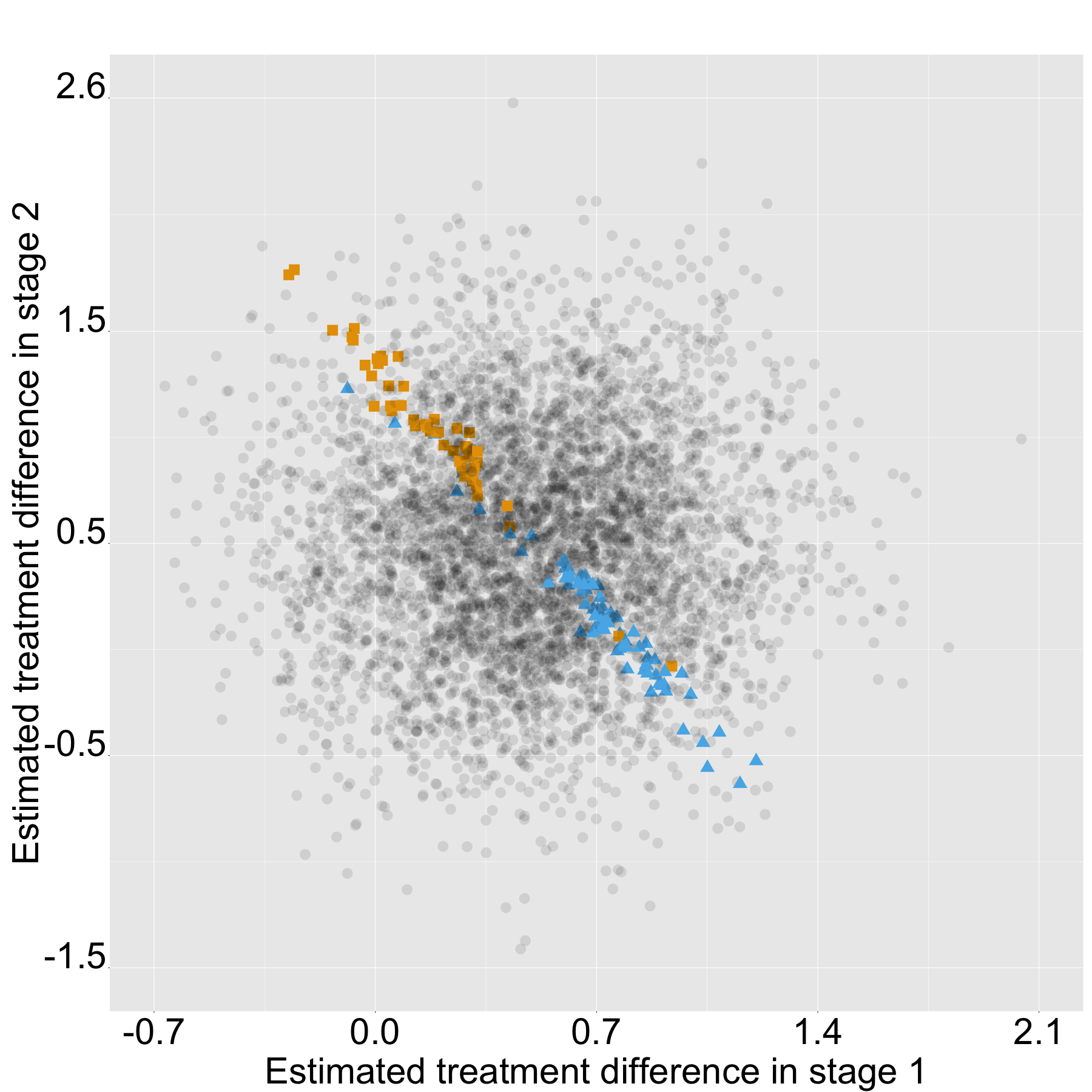}
		\caption{}
		\label{fig:sim_diff_1} 
	\end{subfigure}
	
	\begin{subfigure}[b]{0.55\textwidth}
		\includegraphics[width=1\linewidth]{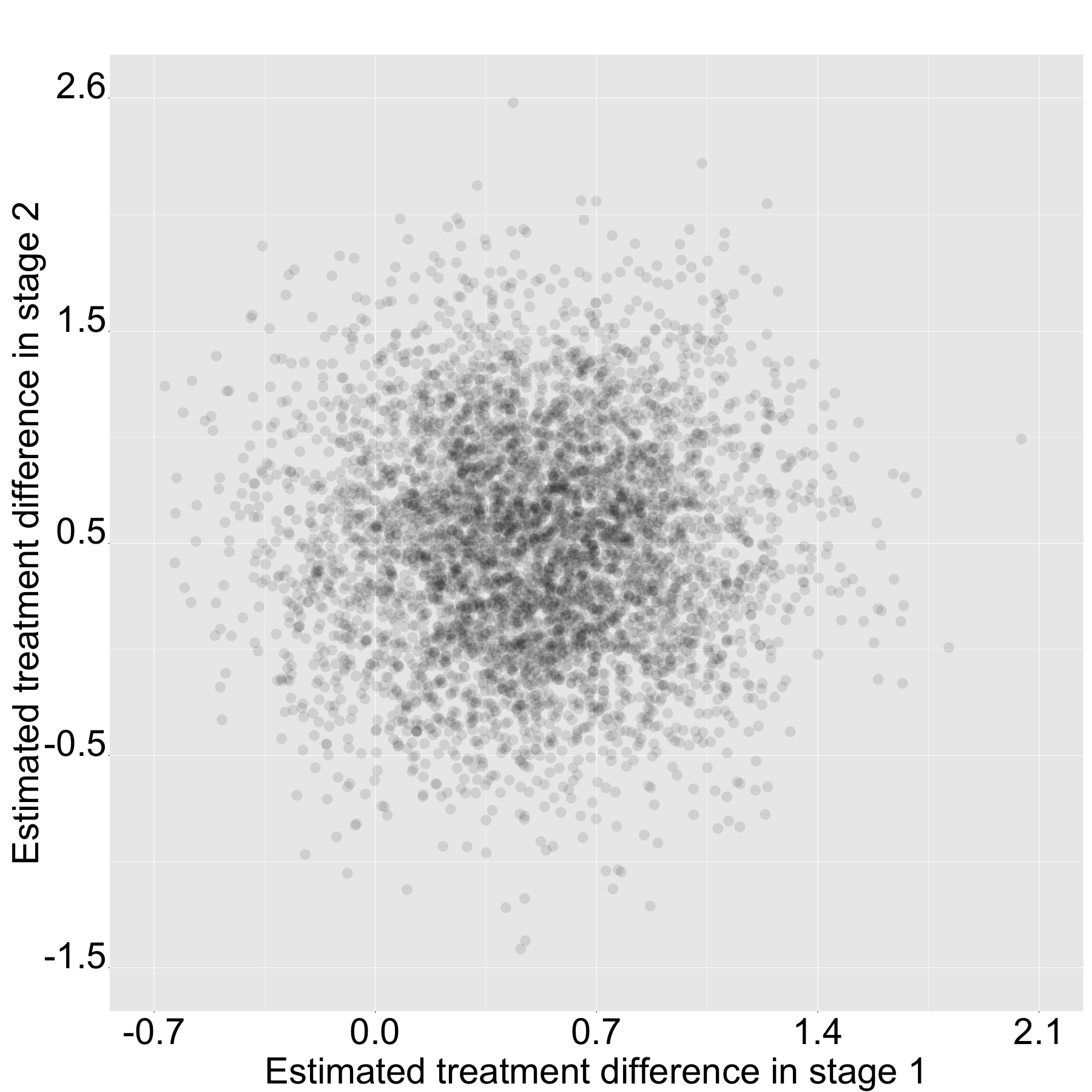}
		\caption{}
		\label{fig:sim_diff_2}
	\end{subfigure}
	\caption[TBD]{Within each figure, the x-axis $\widehat{z}_1$ is the estimated treatment effect based on data from the first stage, and the y-axis $\widehat{z}_2$ is the estimated treatment effect based on data from the second stage. Each dot represents a specific realization of data. It is labeled as gray if both the point estimation of $\theta$ and the hypothesis testing based on $\theta$ have consistent conclusions, but is orange square if only the finding of the point estimation is positive (i.e., the lower bound of the one-sided $1-\alpha$ confidence interval is larger than zero), and is blue triangle if only the hypothesis testing is significant (i.e., the p-value is less than $\alpha$). Figure \ref{fig:sim_diff_1} has a noticeable portion of blue or orange dots, indicating inconsistent findings of the combination testing approach for hypothesis testing and the weighted estimator for point estimation. Our framework is coherent based on Figure \ref{fig:sim_diff_2} with only gray dots.}

	\label{fig:sim_diff}
\end{figure}

Next, we demonstrate another advantage of our weighted framework in Section \ref{sec:weight} with consistent conclusions in point estimation of $\theta$ and hypothesis testing based on $\theta$. An alternative approach for hypothesis testing is the combination testing approach that is commonly used in adaptive clinical trials.\citep{bauer1994evaluation, cui1999modification, bretz2009adaptive} In Figure \ref{fig:sim_diff_1}, we use $\widehat{\theta}_w(\widetilde{w})$ in (\ref{equ:est_w}) with a weight of $\widetilde{w}$ for point estimation, and the combination testing approach with a weight of $\sqrt{\widetilde{w}}$ for the first stage (to ensure that the sum of squares of weights is one) for hypothesis testing. {The combination testing approach rejects the null hypothesis if the combined p-value} $p_{\text{com}} = 1 - \Phi\left[ \sqrt{\widetilde{w}} \Phi^{-1}(1-p_1) + \sqrt{1-\widetilde{w}} \Phi^{-1}(1-p_2) \right]$ {is less than the significance level} $\alpha$, where $p_1$ and $p_2$ {are stage-wise p-values comparing group T versus P in stage 1 and stage 2, respectively, and} $\Phi(\cdot)$ {is the cumulative distribution function of the standard Normal distribution.} Figure \ref{fig:sim_diff_2} reports results of our framework with $\widehat{\theta}_w(\widetilde{w})$ for point estimation and $\widehat{z}_w(\widetilde{w})$ in (\ref{equ:weight_test}) for hypothesis testing. 

In Figures \ref{fig:sim_diff_1} and \ref{fig:sim_diff_2}, the x-axis $\widehat{z}_1$ is the estimated treatment effect based on data from the first stage, and the y-axis $\widehat{z}_2$ is the estimated treatment effect based on data from the second stage. Each dot represents a specific realization of data, with a total of $5,000$ iterations under S5 for visualization. It is labeled as gray if both the point estimation of $\theta$ and the hypothesis testing based on $\theta$ have consistent conclusions, but is orange square if only the point estimation has a positive finding (i.e., the lower bound of the one-sided $1-\alpha$ confidence interval is larger than zero), and is blue triangle if only the hypothesis testing is significant (i.e., the p-value is less than $\alpha$). Figure \ref{fig:sim_diff_1} has a noticeable portion of blue or orange dots, indicating inconsistent findings of the combination testing approach for hypothesis testing and the weighted estimator for point estimation. Our framework is coherent based on Figure \ref{fig:sim_diff_2} with only gray dots.

In conclusion, several weighted estimators in Section \ref{sec:weight} (including the IPTW estimator $\widehat{\theta}_{IP}$ in Section \ref{sec:IPTW} as a special case) and WLS estimators can accurately estimate the treatment effect $\theta$ and control type I error rates under all scenarios considered, especially with different placebo responses and randomization ratios from two stages. The estimator $\widehat{\theta}_w(\widehat{w})$ has very similar power and MSE performance {with two oracle estimators} $\widehat{\theta}_w(w_{opt})$ and $\widehat{\theta}_{\text{WLS}}({\sigma})$, {and the practical WLS estimator} $\widehat{\theta}_{\text{WLS}}(\widehat{\sigma})$. {The proposed} $\widehat{\theta}_w(\widehat{w})$ {is more computationally efficient than} $\widehat{\theta}_{\text{WLS}}(\widehat{\sigma})$. Additionally, our framework in Section \ref{sec:weight} delivers consistent findings for point estimation and hypothesis testing, as compared with an alternative approach of directly using the combination testing approach in adaptive clinical trials for hypothesis testing. 

\begin{table}[ht]
	\centering
	\footnotesize
	\begin{tabular}{cccccc|c|c|cc}
		\cmidrule{1-10}  
		& &   \multicolumn{8}{c }{Bias $\times$ 100 (MSE $\times$ 100)}        \\
		\cmidrule{3-10}                    
		Scenario & $\theta$ & $\widehat{\theta}_d$ &  $\widehat{\theta}_{IP}$ & $\widehat{\theta}_w(\widetilde{w})$ &
		$\widehat{\theta}_w(\widehat{w})$ & $\widehat{\theta}_w(w_{opt})$ & $\widehat{\theta}_{\text{WLS}}({\sigma})$ & $\widehat{\theta}_{\text{LS}}$ & $\widehat{\theta}_{\text{WLS}}(\widehat{\sigma})$ \\
		\cmidrule{1-10}  
S1 & 0 & 0.00 (3.33) & 0.00 (3.33) & 0.00 (3.33) & 0.00 (3.35) & 0.00 (3.33) & 0.00 (3.33) & 0.00 (3.33) & 0.00 (3.35) \\
S2 &  & 0.00 (3.89) & 0.00 (4.01) & 0.00 (4.00) & 0.00 (4.02) & 0.00 (4.00) & 0.00 (4.00) & 0.00 (4.00) & 0.00 (4.02) \\
S3 &  & 0.02 (3.33) & 0.02 (3.33) & 0.02 (3.33) & 0.02 (3.35) & 0.02 (3.33) & 0.02 (3.33) & 0.02 (3.33) & 0.02 (3.35) \\
S4 &  & -4.97 (4.13) & 0.03 (4.00) & 0.03 (3.99) & 0.03 (4.02) & 0.03 (3.99) & 0.03 (3.99) & 0.03 (3.99) & 0.03 (4.01) \\
S5 &  & -4.96 (9.55) & 0.03 (9.66) & 0.03 (9.49) & 0.04 (9.44) & 0.04 (9.34) & 0.04 (9.34) & 0.03 (9.49) & 0.04 (9.44) \\
S6 &  & -5.02 (7.47) & -0.01 (7.38) & -0.01 (7.50) & -0.02 (7.33) & -0.02 (7.28) & -0.02 (7.28) & -0.01 (7.50) & -0.02 (7.33) \\
S7 &  & -5.01 (4.44) & -0.02 (4.72) & -0.02 (4.43) & -0.01 (3.40) & -0.01 (3.38) & -0.01 (3.38) & -0.02 (4.43) & -0.01 (3.40) \\
\\
S1 & 0.5 & 0.01 (3.33) & 0.01 (3.33) & 0.01 (3.33) & 0.01 (3.35) & 0.01 (3.33) & 0.01 (3.33) & 0.01 (3.33) & 0.01 (3.35) \\
S2 &  & -0.03 (3.89) & -0.03 (4.01) & -0.03 (4.00) & -0.03 (4.02) & -0.03 (4.00) & -0.03 (4.00) & -0.03 (4.00) & -0.03 (4.02) \\
S3 &  & -0.01 (3.33) & -0.01 (3.33) & -0.01 (3.33) & -0.01 (3.34) & -0.01 (3.33) & -0.01 (3.33) & -0.01 (3.33) & -0.01 (3.34) \\
S4 &  & -5.01 (4.14) & -0.01 (4.01) & -0.01 (4.00) & -0.01 (4.02) & -0.01 (4.00) & -0.01 (4.00) & -0.01 (4.00) & -0.01 (4.02) \\
S5 & 0.6 & -5.01 (9.56) & -0.02 (9.68) & -0.02 (9.50) & -0.01 (9.45) & -0.01 (9.36) & -0.01 (9.36) & -0.02 (9.50) & -0.01 (9.45) \\
S6 & 0.7 & -5.01 (7.48) & -0.01 (7.39) & -0.01 (7.51) & -0.01 (7.35) & -0.01 (7.29) & -0.01 (7.29) & -0.01 (7.51) & -0.01 (7.35) \\
S7 & 0.5 & -5.01 (4.45) & -0.01 (4.74) & -0.01 (4.44) & 0.00 (3.42) & 0.00 (3.40) & 0.00 (3.40) & -0.01 (4.44) & 0.00 (3.42) \\
		\hline
	\end{tabular}
	\caption{The bias $\times$ 100 and MSE $\times$ 100 in parenthesis of estimators: a direct estimator $\widehat{\theta}_d$ in (\ref{equ:direct}), $\widehat{\theta}_{IP}$ in (\ref{equ:IPTW}) or equivalently $\widehat{\theta}_w\left[(n_{1, \text{T}} + n_{1, \text{P}})/(n_{1, \text{P}} + n_{1, \text{T}} + n_{2, \text{P}} + n_{2, \text{T}})\right]$, $\widehat{\theta}_w(\widetilde{w})$ in (\ref{equ:est_w}) with the optimal weight $w_{opt}$ in (\ref{equ:w_opt_two}) calculated by an assumption of equal variance of two groups, $\widehat{\theta}_w(\widehat{w})$ with $\widehat{w}$ calculated by data to approximate $w_{opt}$, an oracle estimator $\widehat{\theta}_w(w_{opt})$ with the unknown optimal weight $w_{opt}$, an oracle WLS estimator $\widehat{\theta}_{\text{WLS}}({\sigma})$ with unknown variance in Section \ref{sec:wls_oracle}, the least squares estimator $\widehat{\theta}_{\text{LS}}$, and the WLS estimator $\widehat{\theta}_{\text{WLS}}(\widehat{\sigma})$ with estimated variance in Section \ref{sec:wls_est}.}
	\label{tab:sim_1_bias}
\end{table}

\begin{table}[ht]
	\centering
	\begin{tabular}{cccccc|c|c|cc}
		\cmidrule{1-10}   
		& & \multicolumn{8}{c }{Type I error rate}        \\
		\cmidrule{3-10}    
		Scenario & $\theta$ & $\widehat{\theta}_d$ &  $\widehat{\theta}_{IP}$ & $\widehat{\theta}_w(\widetilde{w})$ &
		$\widehat{\theta}_w(\widehat{w})$ & $\widehat{\theta}_w(w_{opt})$ & $\widehat{\theta}_{\text{WLS}}({\sigma})$ & $\widehat{\theta}_{\text{LS}}$ & $\widehat{\theta}_{\text{WLS}}(\widehat{\sigma})$ \\
		\cmidrule{1-10}  
S1 & 0 & 5.03\% & 5.03\% & 5.03\% & 5.11\% & 5.03\% & 5.00\% & 5.00\% & 5.07\% \\
S2 &  & 5.05\% & 5.05\% & 5.05\% & 5.17\% & 5.05\% & 5.00\% & 5.00\% & 5.13\% \\
S3 &  & 4.98\% & 5.03\% & 5.03\% & 5.10\% & 5.03\% & 5.00\% & 5.00\% & 5.06\% \\
S4 &  & 2.88\% & 5.07\% & 5.07\% & 5.17\% & 5.07\% & 5.02\% & 5.02\% & 5.13\% \\
S5 &  & 3.56\% & 5.06\% & 5.07\% & 5.26\% & 5.08\% & 5.00\% & 7.31\% & 5.30\% \\
S6 &  & 3.36\% & 5.04\% & 5.05\% & 5.16\% & 5.03\% & 4.97\% & 3.15\% & 5.17\% \\
S7 &  & 2.89\% & 5.04\% & 5.02\% & 5.13\% & 5.04\% & 4.98\% & 6.19\% & 5.10\% \\
		\cmidrule{1-10}  
		\\
		\cmidrule{1-10}  
		& &  \multicolumn{8}{c }{Power}        \\
		\cmidrule{3-10}    
		Scenario & $\theta$ & $\widehat{\theta}_d$ &  $\widehat{\theta}_{IP}$ & $\widehat{\theta}_w(\widetilde{w})$ &
		$\widehat{\theta}_w(\widehat{w})$ & $\widehat{\theta}_w(w_{opt})$ & $\widehat{\theta}_{\text{WLS}}({\sigma})$ & $\widehat{\theta}_{\text{LS}}$ & $\widehat{\theta}_{\text{WLS}}(\widehat{\sigma})$  \\
		\cmidrule{1-10}  
  S1 & 0.5 & 86.29\% & 86.29\% & 86.29\% & 86.31\% & 86.29\% & 86.22\% & 86.22\% & 86.24\% \\
  S2 &  & 81.31\% & 80.21\% & 80.33\% & 80.40\% & 80.33\% & 80.24\% & 80.24\% & 80.31\% \\
  S3 &  & 86.20\% & 86.30\% & 86.30\% & 86.32\% & 86.30\% & 86.23\% & 86.23\% & 86.25\% \\
  S4 &  & 73.66\% & 80.22\% & 80.35\% & 80.43\% & 80.35\% & 80.26\% & 80.26\% & 80.34\% \\
  S5 & 0.6 & 56.24\% & 61.20\% & 61.86\% & 62.72\% & 62.47\% & 62.31\% &- & 62.86\% \\
  S6 & 0.7 & 77.92\% & 82.36\% & 81.82\% & 82.86\% & 82.78\% & 82.71\% & 75.51\% & 82.89\% \\
  S7 & 0.5 & 70.90\% & 74.34\% & 76.69\% & 85.71\% & 85.67\% & 85.59\% & - & 85.64\% \\
		\hline
	\end{tabular}
	\caption{Type I error rate / power of estimators: a direct estimator $\widehat{\theta}_d$ in (\ref{equ:direct}), $\widehat{\theta}_{IP}$ in (\ref{equ:IPTW}) or equivalently $\widehat{\theta}_w\left[(n_{1, \text{T}} + n_{1, \text{P}})/(n_{1, \text{P}} + n_{1, \text{T}} + n_{2, \text{P}} + n_{2, \text{T}})\right]$ in (\ref{equ:est_w}) with the optimal weight $w_{opt}$ in (\ref{equ:w_opt_two}) calculated by an assumption of equal variance of two groups, $\widehat{\theta}_w(\widehat{w})$ with $\widehat{w}$ calculated by data to approximate $w_{opt}$, an oracle estimator $\widehat{\theta}_w(w_{opt})$ with the unknown optimal weight $w_{opt}$, an oracle WLS estimator $\widehat{\theta}_{\text{WLS}}({\sigma})$ with unknown variance in Section \ref{sec:wls_oracle}, the least squares estimator $\widehat{\theta}_{\text{LS}}$, and the WLS estimator $\widehat{\theta}_{\text{WLS}}(\widehat{\sigma})$ with estimated variance in Section \ref{sec:wls_est}. The power of $\widehat{\theta}_{\text{LS}}$ under S5 and S7 is not shown due to its type I error rate inflation. }
	\label{tab:sim_1_power}
\end{table}

\begin{table}[ht]
	\centering
	\begin{tabular}{ccccc}
		\cmidrule{1-5}   
		& & \multicolumn{3}{c }{Power}       \\
		\cmidrule{3-5}   
		Scenario &  $\theta$ & Planned & 
		$\widehat{\theta}_w({w}_{opt})$ & $\widehat{\theta}_w(\widehat{{w}})$ \\  
		\cmidrule{1-5}    
S1 & 0.5 & 86.30\% & 86.29\% & 86.31\% \\
S2 &  & 80.38\% & 80.33\% & 80.40\% \\
S3 &  & 86.30\% & 86.30\% & 86.32\% \\
S4 &  & 80.38\% & 80.35\% & 80.43\% \\
S5 & 0.6 & 62.45\% & 62.47\% & 62.72\% \\
S6 & 0.7 & 82.86\% & 82.78\% & 82.86\% \\
S7 & 0.5 & 85.74\% & 85.67\% & 85.71\% \\
		\hline
	\end{tabular}
	\caption{Planned power based on ${w}_{opt}$ and empirical power of $\widehat{\theta}_w({w}_{opt})$ and $\widehat{\theta}_w(\widehat{{w}})$.}
	\label{tab:sim_1_pow}
\end{table}

\section{A Case Study}
\label{sec:real}

We apply our method to the Accelerating COVID-19 Therapeutic Interventions and Vaccines (ACTIV) platform trial to evaluate treatments and vaccines for COVID-19.\citep{lavange2021accelerating, ross2023learning} We consider 3 substudies within this master protocol with overlapping enrollment: a randomized trial evaluating ivermectin 400 $\mu$g/kg daily for 3 days versus placebo with enrollment from Jun 2021 to Feb 2022 \citep{naggie2022effect}, a randomized trial of fluvoxamine 50 mg twice daily for 10 days versus placebo with enrollment from Aug 2021 to May 2022 \citep{mccarthy2023effect}, and a randomized trial evaluating ivermectin 600 $\mu$g/kg daily for 6 days versus placebo with enrollment from Feb 2022 to Jul 2022.\citep{naggie2023effect} We denote study drugs in three substudies as Treatment B, Treatment C, and Treatment D, respectively, and their names as Substudy B, Substudy C, and Substudy D, respectively. 

\begin{figure}[h]
	\centering
	\includegraphics[width=16cm]{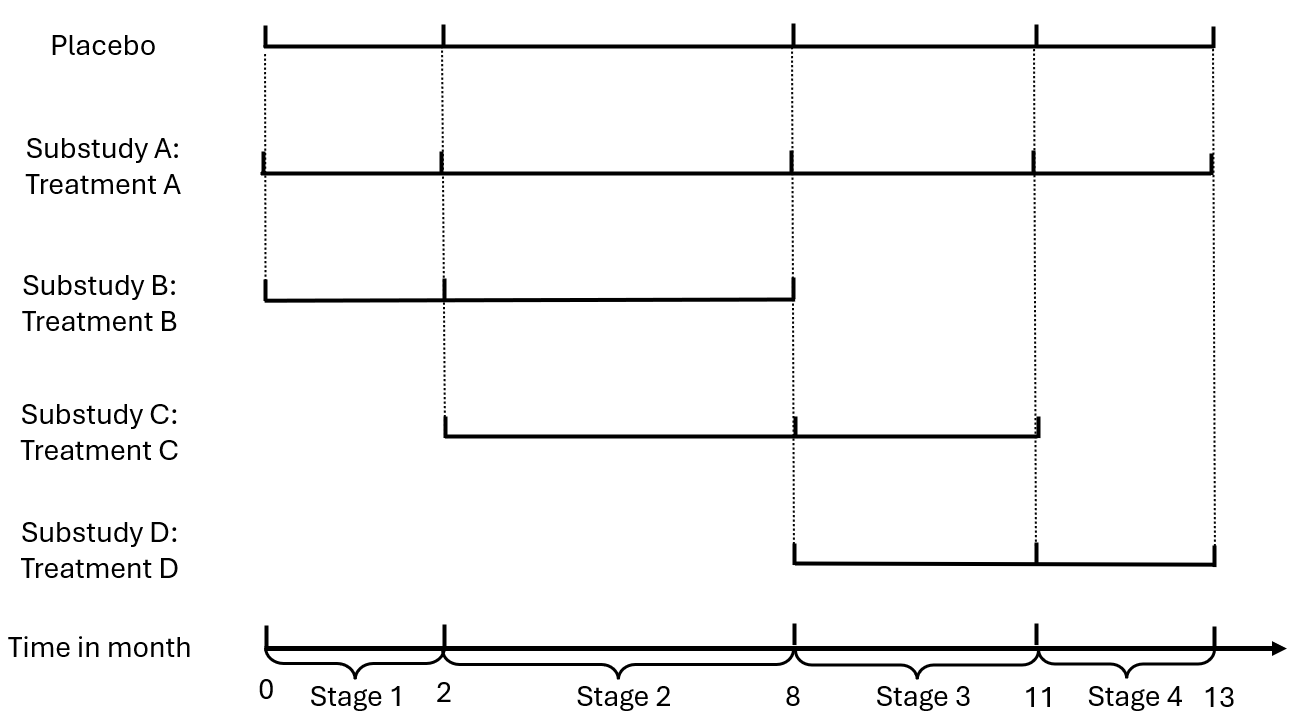}
	\caption{A diagram of Substudy A as the main interest, and three other substudies within the same platform trial.}
	\label{fig:real}
\end{figure}

In this example, we consider another hypothetical Substudy A evaluating Treatment A versus placebo within this platform trial with enrollment from Jun 2021 to Jul 2022. As shown in the study diagram in Figure \ref{fig:real}, the x-axis of time starts with Jun 2021 as zero and has a unit of the month. Based on the enrollment start and end months of 3 existing substudies, there are 4 stages with time lengths of 2, 6, 3, and 2 months, respectively. Suppose that there are a total of $N = 1300$ patients to be uniformly enrolled in this platform trial within the time frame of 13 months, i.e., $100$ patients per month. We assume that all patients are eligible to receive study drugs that are under investigation at the time of enrollment. For example, patients enrolled in Stage 1 are able to receive Treatment A, Treatment B and placebo. Following an example in the FDA guidance \citep{fda2023}, we consider $N_s / \left[\sqrt{k_s}(1+\sqrt{k_s}) \right]$ to be randomized to Treatment A and $N_s / \left({1+\sqrt{k_s}} \right)$ to placebo (concurrent control), with a randomization ratio $r_s$ of Treatment A versus placebo in Stage $s$ as $1/\sqrt{k_s}$, where $k_s$ is the number of active study drugs being evaluated at this stage $s$ and $N_s$ is the number of patients in stage $s$, for $s = 1, 2, 3, 4$. Therefore, we have $r_1 = 1/\sqrt{2}$, $r_2 = 1/\sqrt{3}$, $r_3 = 1/\sqrt{3}$ and $r_4 = 1/\sqrt{2}$. 

The outcome of interest is the amount of time feeling unwell with COVID-19 in days. Motivated by the results of this continuous endpoint in those three substudies \citep{naggie2022effect, mccarthy2023effect, naggie2023effect}, we consider the following 3 scenarios with varying placebo mean parameters: 
\begin{itemize}
	\item S1: $\mu_1 = 11.08$, $\mu_2 = 10.3$, $\mu_3 = 10.45$, $\mu_4 = 11.6$
	\item S2: $\mu_1 = 11.08$, $\mu_2 = 12.3$, $\mu_3 = 12.45$, $\mu_4 = 11.6$
	\item S3: $\mu_1 = 11.6$, $\mu_2 = 12.45$, $\mu_3 = 10.3$, $\mu_4 = 11.08$
\end{itemize}
{We assume a hypothetical setting with} $\sigma_{1,\text{T}} = 1.4$, $\sigma_{2,\text{T}} = 2.7$, $\sigma_{3,\text{T}} = 2$, $\sigma_{4,\text{T}} = 3.3$, $\sigma_{1,\text{P}} = 2$, $\sigma_{2,\text{P}} = 1.2$, $\sigma_{3,\text{P}} = 3.5$ and $\sigma_{4,\text{P}} = 2.9$. Three estimators are evaluated: a direct estimator $\widehat{\theta}_d$ in (\ref{equ:direct}), the IPTW estimator $\widehat{\theta}_{IP}$ in (\ref{equ:IPTW}), an oracle estimator $\widehat{\theta}_w(\boldsymbol{w}_{opt})$ in (\ref{equ:est_w_multi}) with the unknown optimal weight $\boldsymbol{w}_{opt}$ in (\ref{equ:w_opt_multi}), and $\widehat{\theta}_w(\widehat{\boldsymbol{w}})$ with $\widehat{\boldsymbol{w}}$ as an estimate of $\boldsymbol{w}_{opt}$ based on observed data.

Based on results in Table \ref{tab:sim_2}, those weighted estimators $\widehat{\theta}_{IP}$, $\widehat{\theta}_w(\widehat{\boldsymbol{w}})$ and $\widehat{\theta}_w(\boldsymbol{w}_{opt})$ have much smaller bias of estimating $\theta$ than the traditional estimator $\widehat{\theta}_d$. In terms of type I error rate control, $\widehat{\theta}_d$ does not have accurate control, with either an inflated type I error rate (S1) or a deflated one (S2 and S3) with further power loss. The power of $\widehat{\theta}_d$ under S1 is not shown due to its type I error rate inflation. Type I error rates of weighted estimators are close to the nominal level $\alpha=0.05$. In terms of MSE and power, $\widehat{\theta}_w(\widehat{\boldsymbol{w}})$ is better than $\widehat{\theta}_{IP}$ and has a very similar performance with the oracle estimator $\widehat{\theta}_w(\boldsymbol{w}_{opt})$, showing its practical utility. {Section 3 of the Supplementary Materials conducts additional analyses to evaluate Treatment B, C and D, with concurrent controls from stage 1 and 2, stage 2 and 3, and stage 3 and 4, respectively. }

\begin{table}[ht]
	\centering
	\begin{tabular}{ccccc|c|ccc|c}
		\cmidrule{1-10}  
		& &   \multicolumn{4}{c }{Bias $\times$ 100 (MSE $\times$ 100)}  & \multicolumn{4}{c }{Type I error rate / power}       \\
		\cmidrule{3-10}    
		Scenario &  $\theta$ & $\widehat{\theta}_d$ & $\widehat{\theta}_{IP}$ &   $\widehat{\theta}_w(\widehat{\boldsymbol{w}})$ &
		$\widehat{\theta}_w(\boldsymbol{w}_{opt})$ &  $\widehat{\theta}_d$ & $\widehat{\theta}_{IP}$ & $\widehat{\theta}_w(\widehat{\boldsymbol{w}})$ &
		$\widehat{\theta}_w(\boldsymbol{w}_{opt})$  \\
		\cmidrule{1-10}  
S1 & 0 & 4.77 (3.36) & -0.01 (3.14) & -0.00 (2.72) & -0.00 (2.69) & 7.93\% & 5.01\% & 5.19\% & 5.02\% \\
S2 &  & -4.93 (3.38) & -0.02 (3.15) & -0.00 (2.72) & 0.00 (2.69) & 2.55\% & 5.02\% & 5.20\% & 5.04\% \\
S3 &  & -1.69 (3.17) & -0.02 (3.15) & -0.02 (2.72) & -0.02 (2.69) & 3.26\% & 5.01\% & 5.18\% & 5.04\% \\
\\
S1 & 0.45 & 4.79 (3.37) & 0.01 (3.15) & 0.02 (2.72) & 0.02 (2.69) & - & 81.39\% & 86.43\% & 86.37\% \\
S2 &  & -4.94 (3.38) & -0.03 (3.14) & -0.02 (2.71) & -0.02 (2.69) & 71.98\% & 81.36\% & 86.41\% & 86.37\% \\
S3 &  & -1.71 (3.17) & -0.04 (3.16) & -0.03 (2.72) & -0.03 (2.70) & 76.01\% & 81.28\% & 86.35\% & 86.27\% \\
		\hline
	\end{tabular}
	\caption{For the comparison of Treatment A versus Placebo, this table reports the bias $\times$ 100 and MSE $\times$ 100 in parenthesis, type I error rate / power of four estimators: a direct estimator $\widehat{\theta}_d$ in (\ref{equ:direct}), the IPTW estimator $\widehat{\theta}_{IP}$ in (\ref{equ:IPTW}), $\widehat{\theta}_w(\widehat{\boldsymbol{w}})$ in (\ref{equ:est_w_multi}) with $\widehat{\boldsymbol{w}}$ as an estimate of $\boldsymbol{w}_{opt}$ in (\ref{equ:w_opt_multi}) based on observed data, and an oracle estimator $\widehat{\theta}_w(\boldsymbol{w}_{opt})$ with the unknown optimal weight $\boldsymbol{w}_{opt}$. The power of $\widehat{\theta}_d$ under S1 is not shown due to its type I error rate inflation. }
	\label{tab:sim_2}
\end{table}

\section{Discussion}
\label{sec:dis}

In this work, we propose an analysis framework to deliver consistent conclusions for both point estimation and hypothesis testing to facilitate interpretation of platform trial results with a utilization of concurrent controls. Back to those open questions in Section \ref{sec:intro}, we use some theoretical justifications in Section \ref{sec:weight} and empirical evidence in Section \ref{sec:sim} to show that the weighted estimator $\widehat{\theta}_w(w_1)$ in (\ref{equ:est_w}) (the IPTW estimator $\widehat{\theta}_{IP}$ in (\ref{equ:IPTW}) as a special case) can accurately estimate $\theta$, and their corresponding decision rules for hypothesis testing have asymptotic type I error rate control. We further derive the optimal weight $w_{opt}$ in (\ref{equ:w_opt_two}) to gain efficiency, and provide practical suggestions on approximating this unknown weight with observed data in Section \ref{sec:weight}. {Our weighted estimator is closely related to the WLS estimator in Section} \ref{sec:wls}, {and is more computationally efficient.} Additionally, our framework delivers coherent conclusions for both the point estimation and the hypothesis testing with an illustration in Figure \ref{fig:sim_diff_2}. Such properties make the proposed method appealing in both Phase 2 and Phase 3 platform trials with protection of study integrity, increased efficiency and reliable interpretation. 

We consider a basic setting in which the change of randomization ratio does not depend on observed data.\citep{fda2023} In practice, one can consider more complicated adaptive platform trials with potential modifications based on accumulating outcomes data \citep{lin2017comparison, kaizer2018multi, mu2021bayesian}. Many existing approaches in adaptive designs can be readily applied to this setting with some modifications if needed.\citep{bretz2009adaptive} However, additional investigations are needed to identify a consistent estimator with reduced MSE, and a coherent hypothesis testing strategy. One possible approach is to first construct a consistent and efficient estimator of the treatment effect \citep{zhan2024deep}, and then conduct hypothesis testing based on Bootstrap-related methods.\citep{zhan2022finite}

Our proposed method can be readily applied to other endpoints, such as binary and survival endpoints, when the normality of statistics within stages is to be reasonably assumed. For example, Mantel-Haenszel-related methods are able to be applied within each stage to accommodate stratification factors for binary endpoints. The randomization ratios are considered as given in this article, but they can also be optimized for a certain analysis method to gain power. {In the evaluation of a specific treatment group, we only include concurrent controls to provide more robust conclusions} \citep{fda2023}. {More advanced components are needed to properly accommodate non-concurrent controls. Other future works include the generalization with adaptive features discussed above, the adjustment of stage-specific covariates, time-varying placebo responses, etc.} 

\section*{Acknowledgements}
The authors thank the Editor, the Associate Editor and two reviewers for their constructive comments. This manuscript was supported by AbbVie Inc. AbbVie participated in the review and approval of the content. All authors are employed by AbbVie Inc., and may own AbbVie stock.

\section*{Supplementary Material}

The R code to replicate results in Section \ref{sec:sim} and \ref{sec:real} is available on \url{https://github.com/tian-yu-zhan/Analysis_Platform_Trials}.

\bibliographystyle{Chicago}
\bibliography{In_ref.bib}

%

\end{document}